\documentclass[twocolumn,showpacs,prl,superscriptaddress]{revtex4}
\usepackage{graphicx}
\usepackage{dcolumn}
\usepackage{calc}

\begin{document}
\title{Off-shell effects and consistency of many-body treatments of dense matter}

\author{Boris Krippa}
\affiliation{Theoretical Physics Group, Department of Physics and Astronomy, 
University of Manchester, Manchester M13 9PL, UK}
\affiliation{Department of Physics, UMIST, P.O. Box 88, Manchester M60 1QD, UK}
\author{Michael C. Birse}
\affiliation{Theoretical Physics Group, Department of Physics and Astronomy, 
University of Manchester, Manchester M13 9PL, UK}
\author{Judith A. McGovern}
\affiliation{Theoretical Physics Group, Department of Physics and Astronomy, 
University of Manchester, Manchester M13 9PL, UK}
\author{Niels R. Walet}
\affiliation{Department of Physics, UMIST, P.O. Box 88, Manchester M60 1QD, UK}

\date{\today}


\begin{abstract}
Effective field theory requires all observables to be independent of
the representation used for the quantum field operators. It means that
off-shell properties of the interactions should not lead to any
observable effects. We analyse this issue in the context of many-body
approaches to nuclear matter, where it should be possible to shift the 
contributions of lowest order in  purely off-shell two-body interactions 
into three-body forces. We show that none of the commonly used truncations
of the two-body scattering amplitude such as the ladder, 
Brueckner-Hartree-Fock or parquet approximations respect this requirement.

\end{abstract}
\maketitle 

Nuclear forces cannot yet be derived from QCD, and so we must rely
on phenomenological nucleon-nucleon potentials \cite{Machleidt}. There is a 
variety of these potentials, all of which give similarly good fits to the
available low-energy two-body scattering data but which have different off-shell 
behaviours. In recent years it has become clear that a consistent description of
nuclear systems requires three-body forces for both systems consisting
of a small number of nucleons \cite{fewnuc}, as well as for nuclear matter
\cite{Di}.  
Part of the role of the three-body force is to compensate for the different
off-shell behaviours of the two-body forces. This is as expected because 
the physics should not depend on the off-shell behaviour of the interactions. 
Exact calculations with phase-equivalent two-body forces and their corresponding 
three- (and higher-) body forces should therefore give identical results. 
However this leaves open the question of whether the approximate many-body 
techniques used in actual calculations respect this property.

The most elegant way to analyse this problem is to use
effective field theory (EFT), which has become an increasingly popular tool
in modern nuclear physics.  This approach makes use of the fact that
the dynamics at low energy is only weakly dependent on the high-energy
degrees of freedom, and so a detailed knowledge of the interaction at
short distances is not required. The low-energy physics can then be 
described using a local effective Lagrangian 
\cite{We79}. The physical amplitudes can be obtained 
from this in the form of expansions in powers of the low-energy
scales involved.  

Application of the EFT approach to the nucleon-nucleon (NN) system
turns out to be rather complicated due to the large $s$-wave
scattering length, which gives rise to an additional small scale in
the problem.  If the scattering length were  similar to the range of
the NN interaction (usually referred to as ``natural'') then it would
be possible to use a perturbative treatment. Instead, the unnaturally
large scattering length means that the leading terms in the chiral
expansion of the NN potential should be iterated to all orders, as
shown by Weinberg \cite{We91} and 
van Kolck \cite{VK}.  The remaining terms in the potential can be
treated as perturbations, organised according to the power counting
elucidated by Kaplan, Savage and Wise (KSW) \cite{Ka}. 
This can also
be thought of as an expansion around a renormalization-group fixed
point which corresponds to a bound state at threshold \cite{Bi}.

Since the off-shell properties are determined by the representation
chosen for the field operator in EFT's, independence of physics on
off-shell properties is called ``reparametrization invariance".  In
formal field theory the corresponding result is known as the
equivalence theorem \cite{ET}. This issue has been considered in
several recent papers \cite{Fea,Fu1,Fu2}. (Further references can be
found in Ref.~\cite{Fu1}.)  Furnstahl {\it et al.}~\cite{Fu1} have
demonstrated that for a model consisting of a dilute Fermi system with
a \emph{natural} two-body scattering length three-body counter\-terms
can indeed cancel the effects of off-shell part of the two-body
interaction.

Here we explore the extension of these ideas to dense systems with an 
unnatural scattering length, such as nuclear matter. 
We find that the effects of off-shell two-body
interactions can indeed be cancelled by three-body forces. This is
unsurprising given the general nature of the equivalence theorem. Of
more importance are the sets of diagrams that need to be included to
obtain this cancellation, and the implications for commonly used
many-body 
truncation schemes. We shall use our analysis to show inconsistency of
various many-body methods.

At low enough energies we can work with EFT's for the NN system where the 
interaction can be treated as purely short-ranged. In principle these 
should be improved by by including one-pion exchange explicitly, but there 
is still some debate about the best way to do this. (See, for example, 
Ref.~\cite{Be} and references therein.) We consider first an EFT which 
leads to a purely energy-dependent NN potential. To next-to-leading order 
(NLO) in the small scale this spin- and isospin-independent potential has 
the form
\begin{equation}
V_1= C_0 +   C'_2 p^2 ,
\label{eq:lag1}
\end{equation}
where $p^2/M$ is the relative kinetic energy of the nucleons.  
The LO coupling constant $C_0$ is
of order $Q^{-1}$ (where $Q$ is a generic low-energy scale) in KSW
counting and so it should be treated nonperturbatively \cite{Ka}. The
NLO coupling $C_2^\prime$ is proportional to the effective range, and
is thus of order $Q^0$ and can be treated perturbatively in this
counting scheme. Nonetheless, to show an example of the principles
discussed above, we solve the Lippmann-Schwinger (LS) equation with this
potential to all orders in $C_2'$, and find the vacuum $T$-matrix
\begin{equation}
T_1=\frac{C_{0} + p^2 C'_{2}}{1 +\frac{M}{4\pi}(C_{0} + p^2 C'_{2})(ip
+ \mu)}  .
\label{eq:T1}
\end{equation}
Here we have used a subtractive renormalization procedure \cite{Ka,Ge}.
(All coupling constants here should be understood as renormalized ones which 
depend on 
$\mu$ to ensure that the scattering amplitude is $\mu$-independent.)  

More generally, the effective Lagrangian can also include interactions with
space derivatives of the nucleon fields and this leads to a potential
that depends on momentum as well as energy. The most general  NLO 
potential has the form
\begin{equation}
 V_2 = C_0 + C'_{2} p^2 + \frac{1}{2}C_2 ({\bf k}^2+{\bf k}^{\prime 2}-2p^2),
\label{eq:lag2}
\end{equation}
where ${\bf k}$ and ${\bf k}'$ denote the initial and final 
relative momenta of the nucleons. The coupling $C_2$ describes a
purely off-shell interaction. Solving the LS equation
we get
\begin{eqnarray}
T_2&=&T_1\biggl[1+\frac{1}{2(C_{0} + p^2 C'_{2})}\biggl(
C_2({\bf k}^2+{\bf k}^{\prime 2}-2p^2)\cr
&&\qquad-\frac{M}{8\pi}
C_2^2(p^2-{\bf k}^2)(p^2-{\bf k}^{\prime 2})(ip + \mu)
\biggr)\biggr],
\label{eq:T2}
\end{eqnarray}
where $T_1$ is given by Eq.~(\ref{eq:T1}). From this we can see that the two 
$T$-matrices coincide on-shell (${\bf k}^2 = {\bf k}'^{2} = p^2$) and so 
the scattering observables are indeed independent of the off-shell 
behaviour of the potential as required by the equivalence theorem.

\begin{figure}
\includegraphics[height=2cm]{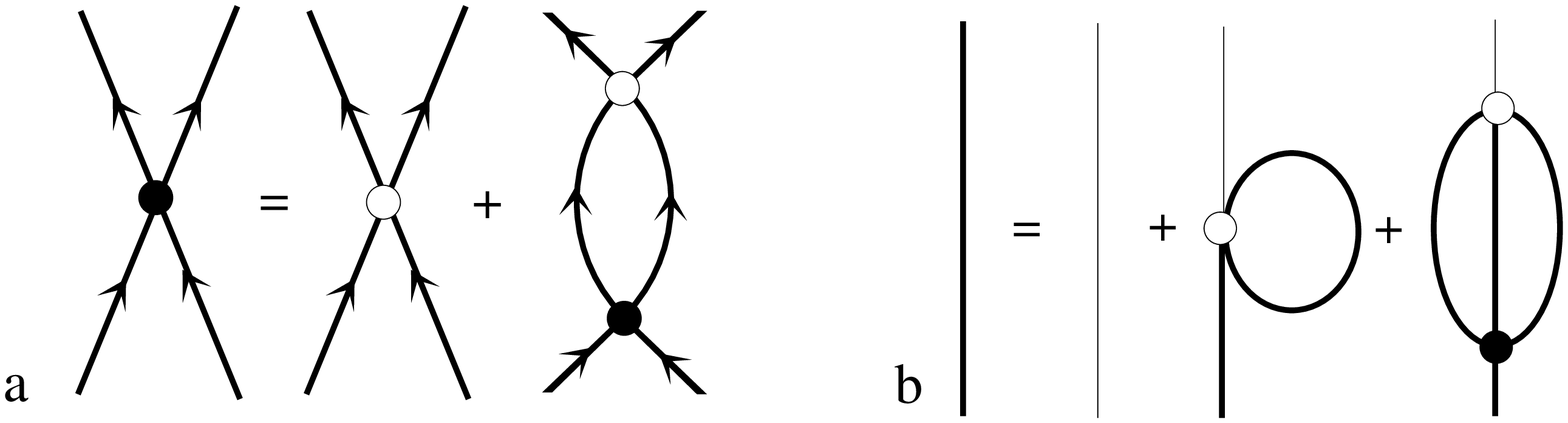}
\caption{Hugenholtz diagrams representing the equation for (a) the $T$ matrix
and (b) the dressed propagator. 
The open circle denotes the LO potential $C_0$, and the solid circle the 
$T$ matrix.}
\label{fig:tm}
\end{figure}

The situation becomes much less trivial in the presence of the nuclear
medium. An in-medium $T$-matrix \cite{Bozek} can be obtained by solving the 
Feynman-Galitskii (FG) equation,
\begin{equation}
T^m = V + VG^{F}T^m,
\label{eq:FG}
\end{equation}
where $G^F$ denotes the two-nucleon propagator which contains  both 
 particle-particle ($pp$) and hole-hole ($hh$) states. This 
$T$-matrix can be thought of as an extension of the more familiar 
$G$-matrix \cite{BR,NO} to include $hh$ as well as $pp$ ladders. It is 
convenient to represent the FG equation graphically in terms of the 
Hugenholtz diagrams of Fig.~\ref{fig:tm}(a). These are versions of Feynman 
diagrams which explicitly incorporate antisymmetry of the interactions. 
Internal lines represent  Feynman propagators which 
describe both particles and holes. The arrows represent the flow of 
quantum numbers such as baryon number. Each topologically distinct 
diagram should be multiplied by a symmetry factor to take account of the 
number of ways it can be constructed from the antisymmetric vertices. 
More details of these diagrams and the rules for evaluating them can be 
found in the textbooks \cite{BR,NO}. 

The solution of the FG equation is rather straightforward in the case of 
zero total momentum of the nucleons. For the potential $V_1$ it takes the 
form
\begin{equation}
T^m_1 = \frac{1}{\frac{1}{T_1}
 + \frac{M}{4\pi^2}[ p\log\frac{p + p_F}{p - p_F}
 - 2p_F]}  ,
\label{eq:Tm1}
\end{equation}
where $p_F$ is the Fermi momentum.
In the same way we can solve the FG equation for the potential
$V_2$. We shall assume that the $C_2$ term
can be treated as a perturbation. 
For simplicity we omit the energy-dependent term 
$C_2^\prime$ from now on. Although this term makes a physically important 
contribution to the energy of the two-particle amplitude, it does not
take part in the cancellation of off-shell effects which is of interest
here. To first order in $C_2$ the in-medium $T$-matrix can be written 
\begin{equation}
T^m_2 = T^m_1 - T^m_1 \frac{C_2}{C_0}(2p^2 - {\bf k}^2 - {\bf k}'^2)
- 2 (T^m_1)^2 \frac{C_2}{C_0}\frac{M}{6\pi^2}p_F^{3}.
\label{eq:Tm2}
\end{equation}

If we now evaluate $T^m_2$ at the on-shell point, we see that it does not 
agree with $T^m_1$ since  the last term does not vanish. This indicates that 
calculations of nuclear matter
based on the in-medium $T$-matrix (or similarly the $G$-matrix) do not 
satisfy the requirement of reparametrization invariance. Alternatively, 
in more traditional nuclear physics language, results for in-medium
observables depend on the off-shell behaviour assumed for the NN
potential.  Such a dependence is unphysical and should not be present. 
A clue to how the dependence may be removed comes from the form of
the final term in Eq.~(\ref{eq:Tm2}), which is proportional to the density.
Its structure is thus similar
to that arising from a three-body contact interaction. This suggests that 
it may be possible to trade off the off-shell dependence against a 
three-body force. As shown below, this can be done, provided our approach 
includes more than just ladder diagrams.

Before exploring what additional physics is needed to remove the
off-shell dependence, we should note that there is no clear separation
of scales in strongly interacting, dense matter. This is an unsolved
problem for the application of EFT's: no power counting has been found
which leads to a consistent expansion. One should really solve the
many-body theory for $C_0$ exactly, by constructing the full in-medium
NN vertex, $\Gamma$. Nonetheless simpler approximations are commonly
used in nuclear physics, typically replacing the full NN vertex by a
$G$- or $T$-matrix. Including $ph$ rings as well as $pp$ and $hh$
ladders leads to the parquet approximation \cite{Par,BR}. We examine
here the consistency of these approximations with reparametrization
invariance. 
\begin{figure}
\hspace*{\fill}\includegraphics[height=1.5cm]{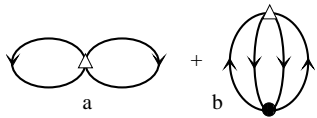}\hspace*{\fill}
\caption{ Hugenholtz diagrams for the ground state energy at first order
order in  $C_2$ (the open triangle).}
\label{fig:GSE}
\end{figure}

\begin{figure}
\hspace*{\fill}\includegraphics[height=2cm]{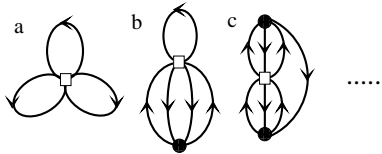}\hspace*{\fill}
\caption{ Hugenholtz diagrams for the ground state energy at first 
order in  the three-body force $D_0$ (the open square).}
\label{fig:3BF}
\end{figure}

At LO in $C_2$ the contributions to the ground-state energy of matter are
shown in Fig.~\ref{fig:GSE}, where the solid dot denotes an in-medium 
NN vertex. If $C_0$ were weak enough we could expand these 
diagrams perturbatively to get a contribution of order $MC_0C_2$.  
The resulting diagrams have an identical structure to those in 
Fig.~\ref{fig:I0} below, except that none of the propagators are dressed. 
As shown in Ref.~\cite{Fu1}, they can be exactly 
cancelled against the LO contribution of a contact three-body interaction 
with strength $D_0=12MC_0C_2$.  This is as required by the equivalence theorem, 
since the off-shell term and three-body force with this strength are both
generated from a Lagrangian which contains neither by the same field redefinition.
The details are given in Ref.~\cite{Fu1}. For definiteness we repeat the relevant
Feynman rules here: the two-body vertices, represented by an open circle and a 
triangle respectively, are
\begin{equation}
 -iC_0 S_2 \quad\hbox{and}\quad
iC_2(\Delta_i+\Delta_{i'}+\Delta_j+\Delta_{j'}) S_2 
\label{eq:2body},
\end{equation}
where $\Delta_i=Mp^0_i-({\bf p}_i)^2/2$, $p_i$ being the four-momentum of the $i$th
nucleon, and the spin-isospin structure is given by 
$S_2=\delta_{ii'}\delta_{jj'}-\delta_{ij'}\delta_{ji'}$. 
The three-body vertex (an open square) is 
\begin{equation}
-iD_0 \Bigl[\delta_{ii'}(\delta_{jj'}\delta_{kk'}-\delta_{jk'}\delta_{kj'})+
\hbox{cyclic}(i',j',k')\Bigr].
\end{equation}
When the leading two-body vertex is resummed we get an effective vertex 
$\Gamma(p_i,p_j;p_{i'},p_{j'}) S_2$ which is denoted by a filled circle
\footnote{Beyond the ladder approximation the NN vertex also has a piece 
which is symmetric in spin-isospin and antisymmetric in momenta.  In the 
integrals we are considering the rest of the momentum dependence is symmetric, 
and so the other structure does not contribute.}.

\begin{figure}
\hspace*{\fill}\includegraphics[height=1.5cm]{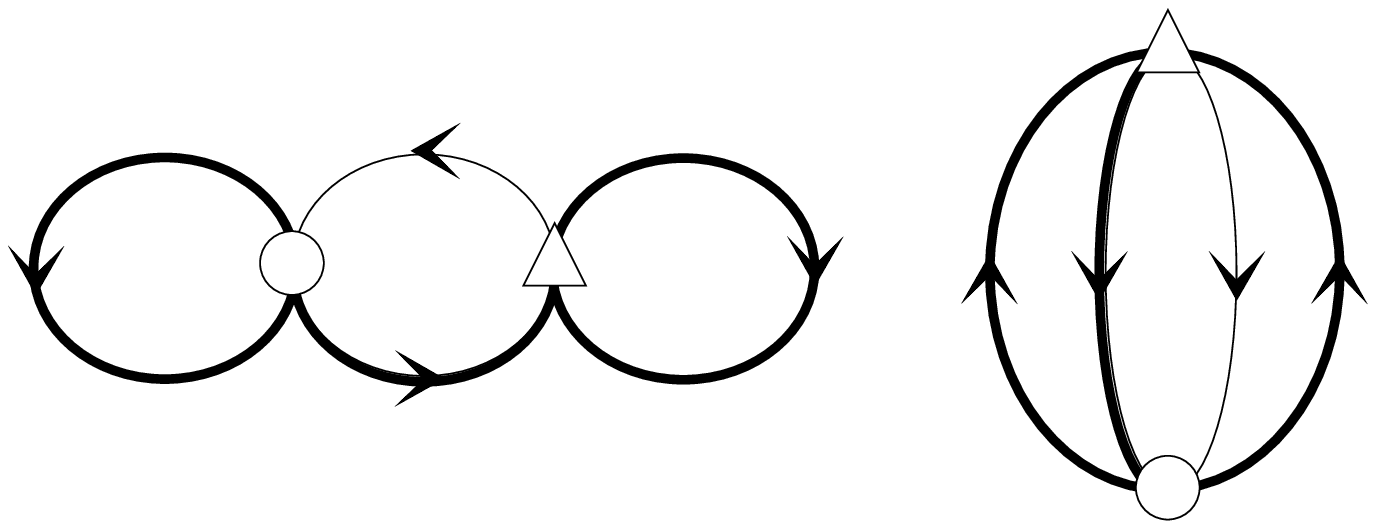}\hspace*{\fill}
\caption{ The contributions from Fig.~\ref{fig:GSE} proportional to the 
integral $I_0$.
Diagrams which can be obtained from those shown by simply reversing
all the arrows are not shown separately.}
\label{fig:I0}
\end{figure}

Treating $C_0$ nonperturbatively, the same three-body force gives rise to 
the diagrams shown in Fig.~\ref{fig:3BF}. Each of these diagrams gives a 
contribution equal to a distinct integral multiplied by $D_0$ and a degeneracy 
factor. The detailed forms of these integrals, which we denote by $I_0$, 
$I_1$, $I_2,\ldots$ are not needed here. However we can use these integrals
to classify the structures which arise from the diagrams of 
Fig.~\ref{fig:GSE}. To evaluate them, we note that the off-shell vertex can 
be written as a sum of four pieces, each of which can cancel a bare 
propagator on one ``leg":
\begin{equation}
 G_0(q)(Mq_0 - {\bf q}^2/2)C_2  = iM C_2,
\label{eq:cancel}
\end{equation}
where $G_0(q)$ is the bare single-particle propagator. The diagrams in 
Fig.~\ref{fig:GSE} give rise to many different contributions, which can be 
identified by iterating the equations for the in-medium NN vertex and 
dressed propagator to pull out a bare propagator ending on a
lowest-order vertex $C_0$ on any of the lines in the original diagrams.
When the bare propagator is cancelled against the off-shell vertex, as in 
Eq.~(\ref{eq:cancel}), the result is one of the integrals $I_n$ multiplied 
by $MC_0C_2$ and a numerical factor. Thus we can
examine the cancellation of off-shell dependence for each integral in turn.

We consider first the Brueckner-Hartree-Fock approximation (BHF) 
\cite{BR,NO}, in which propagators are dressed and the in-medium NN vertex 
is obtained by iterating the potential in the $pp$ and $hh$ channels, as 
shown in Fig.~\ref{fig:tm}. The contributions proportional to $I_0$ from 
Figs.~\ref{fig:GSE}(a) and (b) are shown in Fig.~\ref{fig:I0}.
Except for the dressing of the propagators, these have the same structures
as the perturbative diagrams considered in Ref.~\cite{Fu1} and they can be
shown to cancel with Fig.~\ref{fig:3BF}(a) in the same way.
\begin{figure}
\includegraphics[height=2cm]{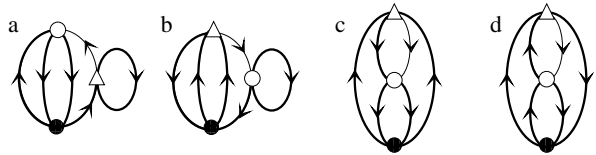}
\caption{(a-c) Contributions proportional to the integral $I_1$ obtained 
from Fig.~\ref{fig:GSE} in the BHF approximation. (d) Extra contribution 
from the parquet approximation. Diagrams which can be obtained from these 
by simply reversing all the arrows are not shown separately.}
\label{fig:I1}
\end{figure}

Fig.~\ref{fig:GSE}(a) gives one other contribution, shown in 
Fig.~\ref{fig:I1}(a),  
which is proportional to $I_1$. In the ladder approximation to the NN vertex, 
Fig.~\ref{fig:GSE}(b) also gives contributions proportional to $I_1$, shown in  
Fig.~\ref{fig:I1}(b-c). The sum of Figs.~\ref{fig:I1}(a-c) is
$-2g(g-1)(2g-3) MC_0 C_2 I_1$, where $g$ is the spin-isospin degeneracy factor 
($g=4$ for symmetric nuclear matter). In contrast, 
the three-body force gives $g(g-1)(g-2)D_0I_1/2$ from Fig.~\ref{fig:3BF}(b). 
We see that the degeneracy coefficients do not agree and cancellation does 
not occur. For example, the off-shell dependence is nonzero for neutron 
matter ($g=2$) where the Pauli principle forbids a contact three-body force.

There is one other structure proportional to $I_1$, Fig.~\ref{fig:I1}(d).
However this cannot be generated from the diagrams of Fig.~\ref{fig:GSE} if 
the potential is iterated in the $pp$ and $hh$ channels only; it requires
iteration in the $ph$ channel as well. When this contribution is included,
the degeneracy factors agree and the off-shell dependence proportional to 
$I_1$ is indeed cancelled by the three-body force with $D_0=12MC_0C_2$.

The crucial point to note is that the cancellation requires diagrams
which can only be obtained by iterating the two-body potential in the 
$ph$ channel.
These are not contained in the ladder or BHF approximations and so 
any approach based on a $G$- or $T$-matrix cannot satisfy the 
equivalence theorem. Observables calculated in these approaches will 
have an unphysical off-shell dependence which cannot be absorbed into 
a three-body force.

The need for diagrams with iteration in the $ph$ channels suggests 
that one should try a more complete 
approach. One such, which treats all two-body channels in a
symmetric way, is the parquet approximation \cite{Par,BR}. If we 
interpret the solid circles in Fig.~\ref{fig:GSE}
as parquet NN vertices constructed from $C_0$, then  all of the
contributions in Fig.~\ref{fig:I1} can be generated by iterating the 
parquet equations. (Note that the parquet self energy can still be 
expressed in the form of Fig.~\ref{fig:tm}(b) \cite{Ja}, and so the discussion 
of $I_0$ above is unchanged.)

\begin{figure}
\includegraphics[height=2cm]{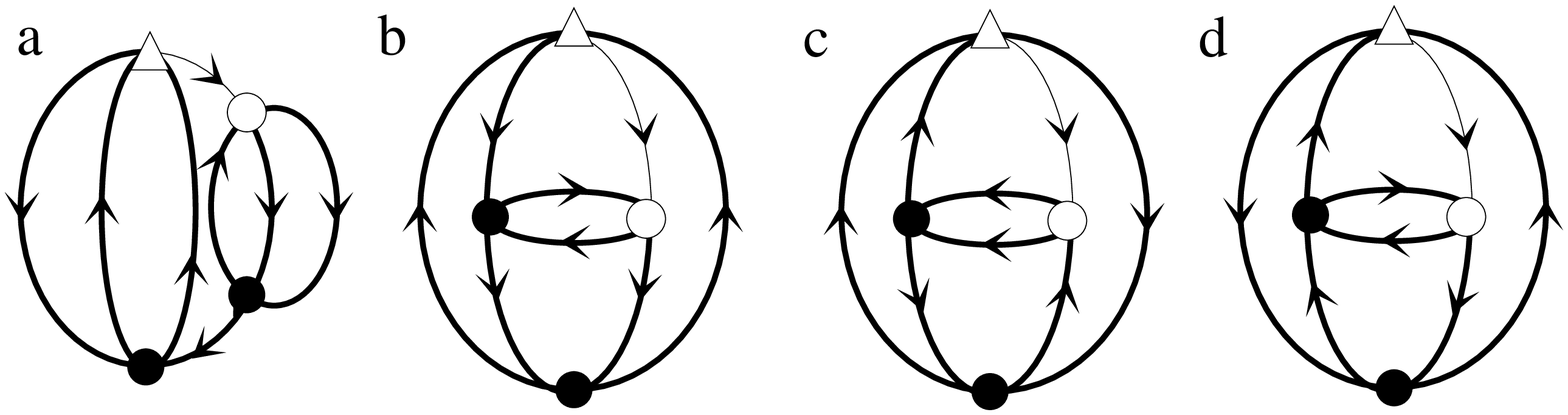}
\caption{(a) Contribution from Fig.~\ref{fig:GSE}(b) proportional to the integral
$I_2$ in the BHF approximation.  (b-c) Extra contribution in the parquet 
approximation. (d) Diagram containing a non-parquet contribution.  
Diagrams which can be obtained from those shown by simply reversing
all the arrows are not shown separately.}
\label{fig:I2}
\end{figure}

Turning now to terms proportional to $I_2$, which ought to cancel with
the three-body graph Fig.~\ref{fig:3BF}(c), we find one self-energy
contribution, Fig.~\ref{fig:I2}(a). This would be present even in the
BHF approximation for the $T$-matrix but, not unexpectedly, this does
not provide the cancellation. In the parquet approximation,
Fig.~\ref{fig:GSE}(b) gives additional contributions, shown in
Fig.~\ref{fig:I2}(b,c), and part of (d). Only when the non-parquet
contributions are included (see Table 3 of Ref.~\cite{Par}), i.e.,
with a full set of diagrams, do we match the result from
Fig.~\ref{fig:3BF}(c).  Thus we conclude that the parquet
approximation also violates reparametrization invariance!

It may well be possible to include the necessary structures by
extending the parquet approximation along the lines discussed in 
Ref.~\cite{BR}, starting from a basic vertex which is a sum
of diagrams which are two-particle irreducible in all channels.
However if, as suggested there, these structures are simply added in 
perturbatively, they will not generate the full in-medium vertices needed 
for the diagrams of Fig.~\ref{fig:I1}.

In summary, our results demonstrate that the requirement of reparametrization
invariance, which would require the effects of off-shell 
dependence of the two-body interaction to be cancelled by a 
three-body force, are not satisfied by any of the commonly used truncations
of the two-body scattering amplitude such as ladder, BHF or parquet 
approximations.  The violations show up in structures with higher numbers 
of insertions
of the in-medium NN vertices for the more sophisticated truncations,
but as the interaction is strong this does not provide a consistent
expansion scheme. Finding such a scheme remains essential for 
practical applications of EFT's to dense, strongly interacting matter. 

This work was supported by the EPSRC.


\begin{thebibliography}{0}
\bibitem{Machleidt} R. Machleidt and I. Slaus,
J.~Phys.~G: Nucl.~Part.~Phys.\ 
\textbf{27} R69 (2001).

\bibitem{fewnuc} S. C. Pieper, V. R. Pandharipande, R. B. Wiringa, and J. Carlson, 
Phys.~Rev.~C {\bf 64},  014001 (2001).

\bibitem{Di} H.Q. Song, M. Baldo, G. Giansiracusa 
 and U. Lombardo, Phys.~Rev.~Lett.\ {\bf 81}, 1584 (1998).

\bibitem{We79} S. Weinberg, Physica A {\bf 96},  32 (1979).

\bibitem{We91} S. Weinberg, Nucl. Phys. {\bf B363},  3 (1991).

\bibitem{VK} U. van Kolck, Nucl. Phys. {\bf A645}, 273 (1999).

\bibitem{Ka} D. B. Kaplan, M. J. Savage and M. B. Wise,
Nucl. Phys. {\bf B534}, 329 (1998).

\bibitem{Bi}   M. C. Birse, J. A. McGovern and K. G. Richardson,  Phys. Lett. {\bf 
B464},  169 (1999).

\bibitem{ET} R. Haag, Phys. Rev. {\bf B112},  (1958) 669;
S. Kamefuchi, L. O'Raifeartaigh and A. Salam, Nucl. Phys. {\bf 28}, 529 (1961).

\bibitem{Fea}  H. Fearing, Phys. Rev. Lett. {\bf 81},  4 (1998); H. Fearing and
S. Scherer, Phys. Rev. {\bf C62}, 054006 (2000).

\bibitem{Fu1} R. J. Furnstahl, H.-W. Hammer and N. Tirfessa,  Nucl. Phys. {\bf A689},
846 (2001).

\bibitem{Fu2} R. J. Furnstahl and H.-W. Hammer, Phys. Lett. {\bf B531}, 203 (2002).

\bibitem{Be} S. R. Beane, P. F. Bedaque, M. J. Savage and U. van Kolck,
Nucl. Phys. {\bf A700}, 377 (2002).

\bibitem{Ge} J. Gegelia,  Phys.~Lett. {\bf B429}, 227 (1998).

\bibitem{Bozek}  P. Bozek,
 Phys.~Rev.~C \textbf{65}, 054306 (2002).

\bibitem{BR} J. P. Blaizot and G. Ripka, {\it Quantum theory of finite systems}
(MIT Press, Cambridge, 1985).

\bibitem{NO} J. W. Negele and H. Orland, {\it Quantum many-particle systems} 
(Perseus, New York, 1988).

\bibitem{Par} A. D. Jackson, A. Lande and R. A. Smith, Phys. Rep. {\bf 86}, 55 
(1982). 

\bibitem{Ja} A. D. Jackson and R. A. Smith, Phys. Rev. {\bf A36}, 2517 (1987).

\end{thebibliography}
\end{document}